\documentclass[12pt,a4paper]{article}

\usepackage{cite}
\usepackage{fleqn,espcrc1}
\usepackage[dvips]{epsfig}
\usepackage{multicol}

\begin{document}
\begin{sloppypar}

\title{Simultaneous Measurements of the $p+d\rightarrow
(A=3)\ +\pi$ Reactions}

\author{GEM Collaboration\\ M. Betigeri$^{\ i}$,
J. Bojowald$^{\ a}$, A. Budzanowski$^{\ d}$, A. Chatterjee$^{\ i}$, J.
Ernst$^{\ g}$, L. Freindl$^{\ d}$, D. Frekers$^{\ h}$, W. Garske$^{\
h}$, K. Grewer$^{\ h}$, A. Hamacher$^{\ a}$, J. Ilieva$^{\ a,e}$, L.
Jarczyk$^{\ c}$, K. Kilian$^{\ a}$, S. Kliczewski$^{\ d}$, W.
Klimala$^{\ a,c}$, D. Kolev$^{\ f}$, T. Kutsarova$^{\ e}$, J. Lieb$^{\
j}$, H. Machner$^{\ a}$\thanks{corresponding author, e-mail:
h.machner@fz-juelich.de}, A. Magiera$^{\ c}$, H. Nann$^{\ a}$\thanks{on
leave from IUCF, Bloomington, Indiana, USA}, L. Pentchev$^{\ e}$, H. S.
Plendl$^{\ k}$, D. Proti\'c$^{\ a}$, B. Razen$^{\ a,g}$, P. von
Rossen$^{\ a}$, B. J. Roy$^{\ i}$, R. Siudak$^{\ d}$, J. Smyrski$^{\
c}$,R. V. Srikantiah$^{\ i}$, A. Strza{\l}kowski$^{\ c}$, R. Tsenov$^{\
f}$, K. Zwoll$^{b}$ \vspace {0.5cm}
\\ \small
\noindent {\it a. Institut f\"{u}r Kernphysik, Forschungszentrum J\"{u}lich, J\"{u}lich, Germany}
\\ \small \noindent {\it b. Zentrallabor f\"ur Elektronik, Forschungszentrum J\"ulich,
J\"ulich, Germany}\\ \small \noindent {\it c. Institute of Physics,
Jagellonian University, Krakow, Poland}
\\ \small \noindent {\it d. Institute of Nuclear Physics, Krakow, Poland}
\\ \small \noindent {\it e. Institute of Nuclear Physics and Nuclear Energy,  Sofia, Bulgaria}
\\ \small \noindent {\it f. Physics Faculty, University of Sofia, Sofia, Bulgaria}
\\ \small \noindent {\it g. Institut f\"ur Strahlen- und Kernphysik der Universit\"at
Bonn, Bonn, Germany}
\\ \small \noindent {\it h. Institut f\"ur Kernphysik,  Universit\"at M\"unster,
M\"unster, Germany}
\\ \small \noindent {\it i. Nuclear Physics Division, BARC, Bombay, India}
\\ \small \noindent {\it j. Physics Department, George Mason University,
Fairfax, Virginia, USA}
\\ \small \noindent {\it k. Physics Department, Florida State University,
Tallahassee, Florida, USA}}

\maketitle
\begin{abstract}
A stack of annular detectors made of high purity germanium was used to
measure simultaneously $pd\rightarrow {^3H}\ \pi^+$ and $pd\rightarrow
{^3He}\ \pi^0$ differential cross sections at beam momenta of 750
MeV/c, 800 MeV/c, and 850 MeV/c over a large angular range. The
extracted total cross sections for the  $pd\rightarrow {^3He}\ \pi^0$
reactions bridge a gap between near threshold data and those in the
resonance region. The ratio of the cross sections for the two reaction
channels taken at the same $\eta={p^{cm}_\pi}/m_\pi$ yields
$2.11\pm0.08$ indicating that a deviation from isospin symmetry is very
small.
\end{abstract}

\section[Introduction]{\label{Introduction}Introduction}

 The
study of the two reactions
\begin{eqnarray}
\label{pd-trit}
  pd\rightarrow {^3H}\ {\pi^+}
\\
  pd\rightarrow {^3He}\ {\pi^0}
  \label{pd-he}
  \end{eqnarray}
is  of interest because of the underlying reaction mechanism. On one
hand, the deuteron is the ideal case for the impulse approximation. In
this approximation the incident proton interacts with one target
nucleon leaving the third nucleon as a spectator. Thus, both reactions
are dominated by the underlying two elementary reactions
\begin{eqnarray}
\label{pp}
  pp\rightarrow d\pi^+
\\
\label{np}
  np\rightarrow d\pi^0.
  \end{eqnarray}
On the other hand, the struck nucleon is bound in the deuteron and its
momentum distribution will strongly influence the reaction since the
presently chosen lowest beam momentum of 750 MeV/c is still below the
pion threshold in nucleon--nucleon interactions.

The model of the underlying elementary reactions \ref{pp} and \ref{np}
goes back to Ruderman \cite{Rud52}. Although many groups have
theoretically studied the two reactions along this path, only moderate
success has been achieved. The various approaches differ in the
treatment of the bound state wave functions, additional interactions,
distortions, inclusion of the deuteron D-state and so on. An excellent
overview of the theoretical work is given by Canton et al.
\cite{Can98}.

Studies of pion production on the nucleon as well as on light nuclei
induced by protons have been shown to be dominated by excitation of the
P-wave $\Delta (1232)$ isobar. Only very close to threshold one is
sensitive to the interesting pion S-wave \cite{Dro98}. While for the
elementary reactions \ref{pp} and \ref{np}, no interference between
these two waves can occur because of the symmetry in the entrance
channel, reactions \ref{pd-trit} and \ref{pd-he} show very strong
forward--backward asymmetry containing information on the phases. The
$NN\to d \pi$ models are therefore especially successful in the
resonance region where the interference is small compared to the P-wave
contribution. Germond and Wilkin \cite{Ger90} showed that for collinear
kinematics only two independent amplitudes exist. These were fitted to
forward and backward scattering angles, to differential cross section
data and to deuteron tensor analyzing powers \cite{Ker86}. It was shown
in Ref. \cite{Ger90} that close to threshold the energy dependence of
these two amplitudes is represented by a linear expansion in terms of
$x=\eta \cos(\theta)$ with $\eta ={p^{cm}_\pi}/m_\pi$ and $\theta$ the
emission angle in the centre of mass system. However, when this model
with the fitted parameters is extended to higher energies, it can
neither describe the total cross sections nor the differential cross
sections at forward and backward angles.
\begin{figure}[htbp]
 \begin{center}
  \epsfig{figure=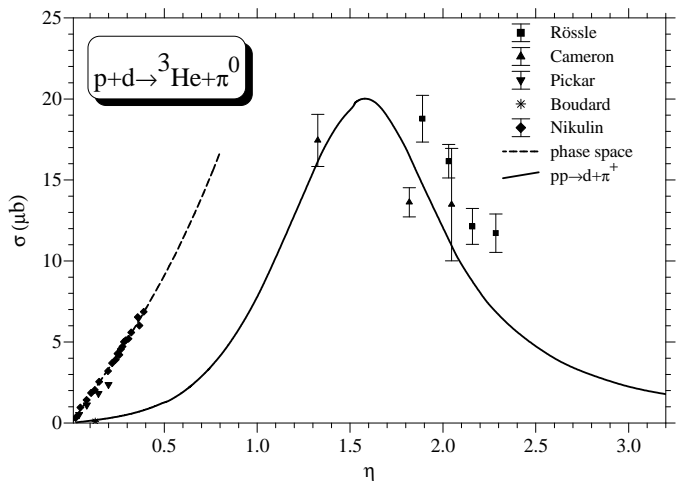,width=12cm,angle=0}
 \end{center}
   \caption{\label{Exfu_intro}Excitation function for the indicated
   reaction. Data are shown as symbols with error bars and are from
   Ref.'s \cite{Bou88,Cam87,Dut81,Nik96,Pic92}. The result of the
   Germond-Wilkin model is shown as dashed curve, and the $pp\to d \pi^+$
   reaction divided by 160 as solid curve.}
\end{figure}
The situation for the total cross section is shown in Fig.
\ref{Exfu_intro}. The data are from Ref.'s
\cite{Bou88,Cam87,Dut81,Nik96,Pic92}. For the data from Ref.
\cite{Dut81}, charge symmetry was assumed to hold. Pickar et al.
\cite{Pic92} extracted total cross sections from the forward and
backward differential cross sections of Kerboul et al. \cite{Ker86} by
making additional use of some systematics. These data are omitted here
since they are much larger than direct measurements. The
near--threshold data with $\eta \leq 0.5$ are reproduced by the Germond
-Wilkin model. The data in the resonance region follow the scaled
$pp\to d\pi^+$ cross section. Clearly, in the intermediate region where
the S--D--wave interference effects are expected to be large, data are
missing. We, therefore, performed measurements for reaction \ref{pd-he}
in that interval. In these experiments, also the production of $pd\to
{^3H}\ \pi^+$ (reaction \ref{pd-trit}) was measured simultaneously. A
comparison of reactions \ref{pd-trit} and \ref{pd-he} permits the study
of isospin symmetry breaking. The beam energy was chosen to be below
the $\Delta $ resonance in order to avoid problems due to different
resonance masses. On the other hand, the momentum seems to be
sufficiently large to minimize Coulomb effects in the exit channel.

Isospin symmetry is known to be only an approximate symmetry. In
addition to static breaking due to Coulomb effects, the dynamic effects
due to the mass differences between the up and down quark are expected
to contribute \cite{Ber98}. Isospin symmetry predicts a value of 2.0
for the ratio of the cross sections. There have been several studies in
the literature of the ratio
\begin{equation}\label{R_ratio}
  R(\theta)=\frac{d\sigma(\theta,pd\rightarrow
^3H\pi^+)/d\Omega}{d\sigma(\theta,pd\rightarrow
^3He\pi^0)/d\Omega}
\end{equation}
for the same beam momentum. However, it is not clear whether this is
the proper quantity to study isospin symmetry conservation because of
the different pion masses and mass differences between $^3H$ and
$^3He$. The great advantage of the present work compared to
measurements with constant pion centre of mass momentum $p_\pi$ is that
a simultaneous measurement avoids a lot of systematical errors. This
principal advantage is lost when the detector does not allow
simultaneous measurements as in Ref.'s \cite{Cre60,Har60,Low81,Sil85}.
Furthermore, these studies are restricted to only a few (up to 4)
angles. All previous results for $R$ are larger than 2. The two
reactions were also measured almost over the full angular range by the
same group with the same apparatus at 500 MeV, however at different
time \cite{Cam81,Cam87}. We found from these measurements an integral
ratio $R=1.82\pm 0.18$. It should be mentioned that the authors never
drew conclusions about charge independence from the cross section data
but studied analyzing powers. K\"{o}hler \cite{Koh60} studied the ratio
$R(\theta)$ in the framework of the Ruderman approach
\cite{Rud52,Fea77} and found the ratio of the form factors to be the
dominant source of deviation from 2.

\section[Experimental Procedure]{\label{Experimental Procedure}Experimental Procedure}

\begin{figure}[htbp]
  \begin{center}
    \epsfig{figure=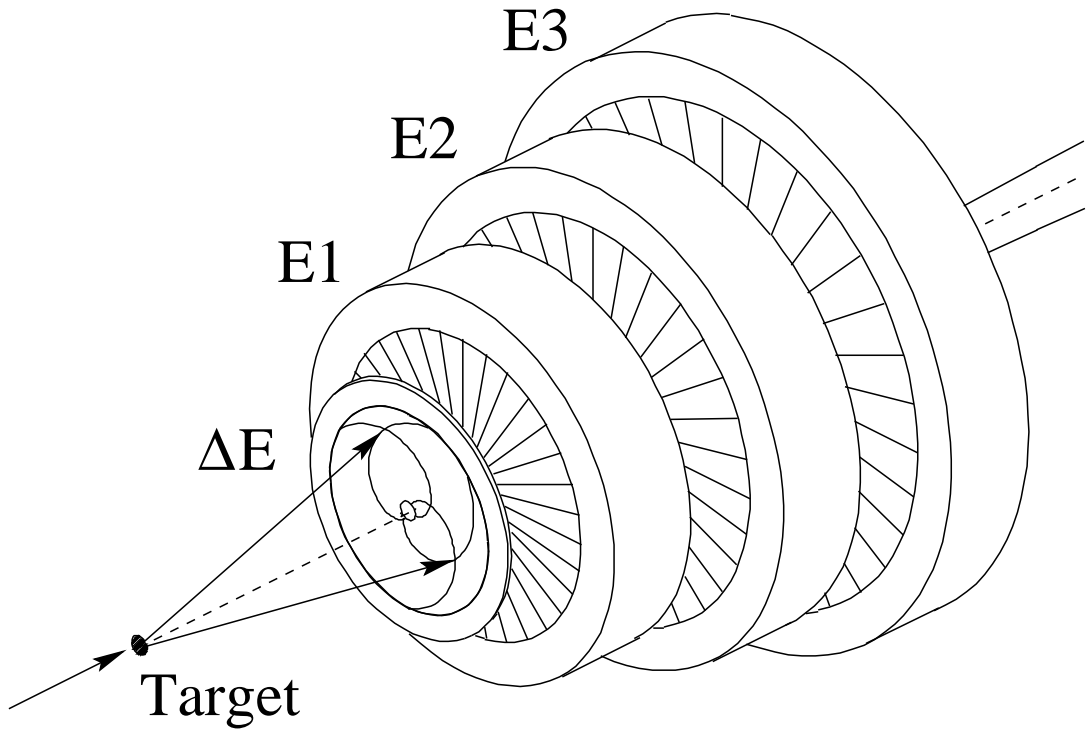,angle=0,width=14.4cm}
     \end{center}
  \caption[Germanium Wall Setup]{\label{GermaniumWall} \small The
  ``Germanium Wall''.}
\end{figure}

We have measured reactions  \ref{pd-trit} and \ref{pd-he}
simultaneously employing a detector with large momentum and geometrical
acceptance for both heavy recoiling nuclei at the same time. Proton
beams with momenta of 750 MeV/c, 800 MeV/c, and 850 MeV/c were
extracted from the COSY accelerator and focussed onto a target cell
containing liquid deuterium. It had a diameter of 6 mm and a thickness
of 6.4 mm \cite{Jae94} with windows of 1.5 $\mu$m Mylar. The excellent
ratio of deuterium to heavier nuclei in the window material reduced
empty target events to a negligible level in contrast to all previous
studies. The beam intensity was measured by different monitor counter
arrangements being individually calibrated for each setting of the
accelerator. The calibration was done by measuring the number of
scattered particles in the monitor counters as a function of the beam
particles. The latter were measured with the trigger hodoscope in the
focal plane of the magnet spectrograph. This number is of course much
larger and leads to dead time in the hodoscope. The beam intensity was
then reduced by debunching the beam between the ion source and the
cyclotron injector. For sufficiently small beam intensity, the relation
between monitors and hodoscope is linear. Because of the chosen
geometry, the counting rate in the monitor counters was small, thus
avoiding errors due to pile up. The detector system is the so called
GEM detector, consisting of the ``{\it GE}rmanium {\it W}all''
\cite{Bet99} and the Q3D2Q {\it M}agnetic spectrograph \cite{Dro98} at
COSY. Here we give only some additional details specific for this
experiment.The germanium wall consisted of three high purity germanium
detectors with radial symmetry with respect to the beam axis as shown
in figure \ref{GermaniumWall}.
The first detector (called Quirl-detector) measures the position and
the energy loss of the penetrating particles. The active area of this
diode is divided on both sides by 200 grooves. Each groove is shaped as
an Archimedes' spiral covering an angular range of $2 \pi$ with
opposite directions on the front and rear side, respectively. The
energy-detectors are mainly used for measuring the energy loss of the
penetrating particles or the total kinetic energy of stopped particles,
respectively. These detectors are divided into 32 wedges to reduce the
counting rate per division leading to a higher maximum total counting
rate of the total detector.

\begin{figure}[htbp]
\begin{center}
\epsfig{file=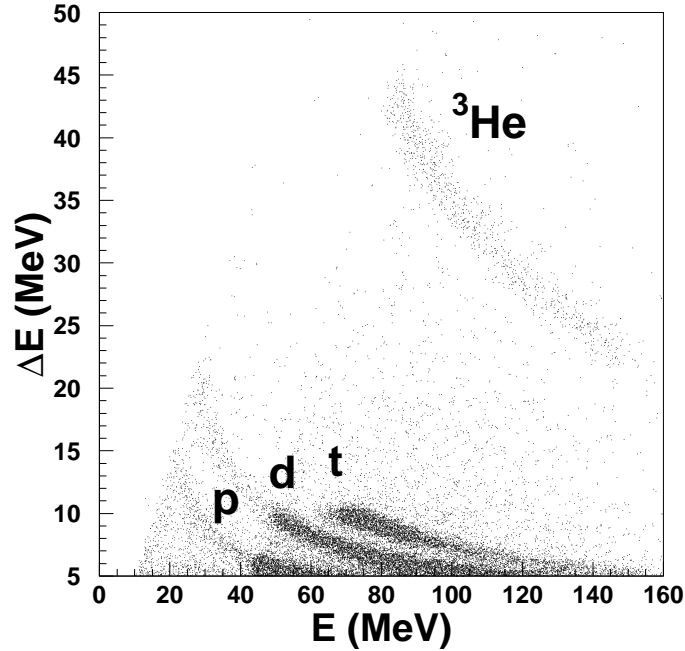,height=10cm}
\end{center}
\caption{\label{Delta_E} \small Energy loss in the Quirl as function of
the total energy deposited in all detectors. The bands visible are due
to detected protons, deuterons, tritons and $^3He$.}
\end{figure}

Fig. \ref{Delta_E} shows the response of the germanium wall for
reaction particles from the interaction of 850 MeV/c protons with
deuterons.
Clearly visible are bands attributed to protons, deuterons, tritons,
and ${^3He}$. The latter two ions are produced in the two reactions of
present interest $pd\rightarrow {^3H}\ \pi^+$ and $pd\rightarrow
{^3He}\ \pi^0$. Protons and deuterons are from elastic scattering and
from $pd\rightarrow pd\pi^0$ reactions. The faint areas for protons and
deuterons are due to the situation where one particle with higher
energy generates a trigger for a low energy particle falling below the
discriminator thresholds. The information deduced from the germanium
wall are energy, emission vertex and particle type. They were converted
to a four momentum vector. These measurements together with the
knowledge of the four momenta in the initial state yield the missing
mass of the unobserved pion by applying conservation of momentum,
energy, charge and baryon number.

\begin{figure}[htbp]
 \begin{center}
  \epsfig{figure=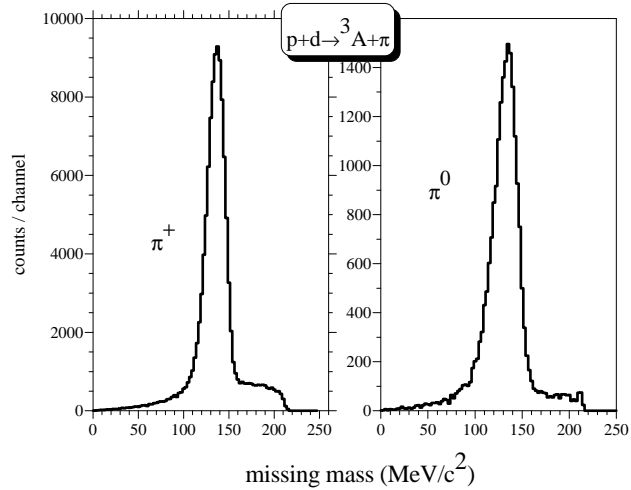, width=10cm, angle=0}
 \end{center}
\caption{\label{missing_mass}Missing mass distributions for the two
reactions at a beam momentum of 850 MeV/c.}
\end{figure}

In the off-line analysis, soft gates were applied to the triton and
$^3He$ loci in Fig. \ref{Delta_E}. This leads to some background in the
missing mass spectrum but avoids throwing away good events. An example
for ${^3H}$ and ${^3He}$ emission at a beam momentum of 850 MeV/c is
shown in Fig. \ref{missing_mass}.
The data are shown as histograms. The good peak to background ratio is
obvious as well as the excellent statistics. The peak in the case of
${^3He}$ emission is broader than for ${^3H}$ emission. This is due to
the larger energy straggeling in the target of the double charged
particle compared to the one for single charged particle. In the final
analysis this effect is corrected. The data could be fitted by a second
order polynomial for the background and a Gaussian with variable width
for the peak. In order to remove this background, spectra of
observables not depending on the emission angle were analyzed. For
${^3He}$ the $\Delta E-E$ curve was linearized and projected to the
$\Delta E$-axis. For ${^3H}$ the centre of mass momentum was used. The
resulting spectra show a peak on a smooth background. In a second step,
the background was fitted by smooth functions and subtracted for each
bin in $\cos (\theta )$. The efficiency of the analysis procedures were
studied by Monte Carlo calculations \cite{Gar99}.

For ${^3He}$ particles being stopped in the Quirl detector, the
kinematical relation between emission angle and energy was applied.
Unfortunately, this was only possible for the runs at 750 MeV/c and 800
MeV/c, while in the runs at 850 MeV/c a hardware coincidence between
the Quirl and the first energy detector was required. Tritons being
emitted under zero degree in the laboratory system with the smallest
energy were detected in the magnetic spectrometer applying hardware and
software cuts as reported in Ref. \cite{Dro98}. Finally, the data were
corrected for reduced efficiencies due to nuclear absorption in the
detector material \cite{Mac99}.

\section[Experimental Results]{\label{Experimental Results}Experimental Results}
\begin{figure}[htbp]
\begin{center}
\epsfig{file=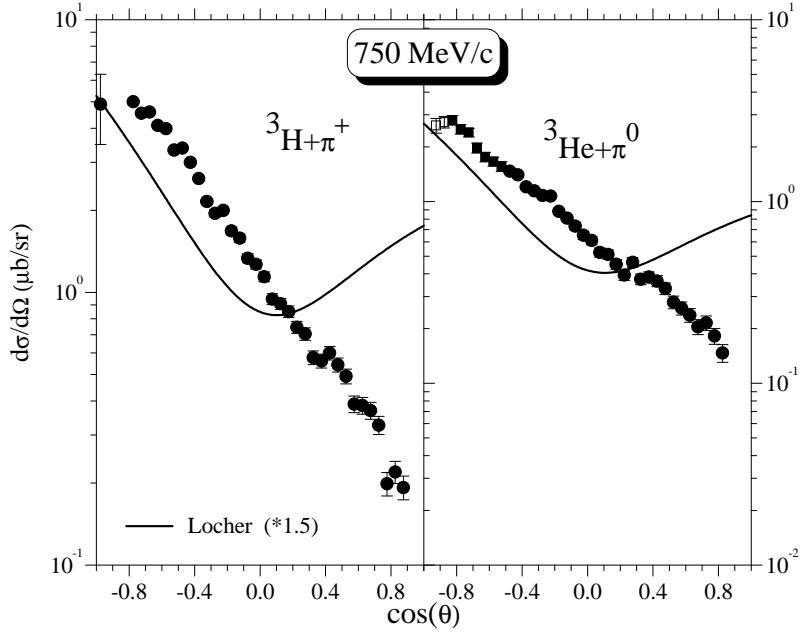,width=14cm} \end{center}
\caption{\label{cross_section_750}Differential cross sections for the
$p+d\to {^3H}+\pi ^+$ reaction at a beam momentum of 750 MeV/c are
shown as full dots. The data are obtained by coincidence measurements
with the germanium wall or by the magnetic spectrograph (only $\cos
(\theta)=-1$). In the case of the $p+d\to {^3He}+\pi ^0$ reaction data
obtained by coincidence measurements are shown as full dots, those
obtained by kinematic relation as full and open squares. The data with
uncertain efficiency because of the vicinity of the corresponding
detector element to the central hole are shown by the open symbol and
are excluded from the further analysis. Calculations in the
Locher--Weber model are shown as solid curves.}
\end{figure}

\begin{figure}[htbp]
\begin{center}
\epsfig{file=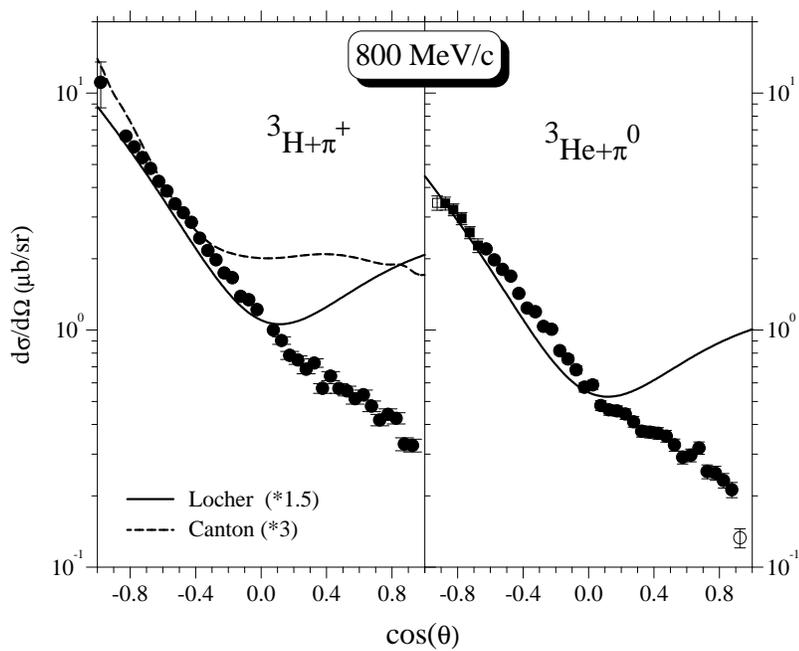,width=14cm} \end{center}
\caption{\label{cross_section_800} Same as Fig. \ref{cross_section_750}
for 800 MeV/c . In addition a calculation from Canton et al.
\cite{Can98} is shown as long dashed curve.}
\end{figure}

\begin{figure}[htbp]
\begin{center}
\epsfig{file=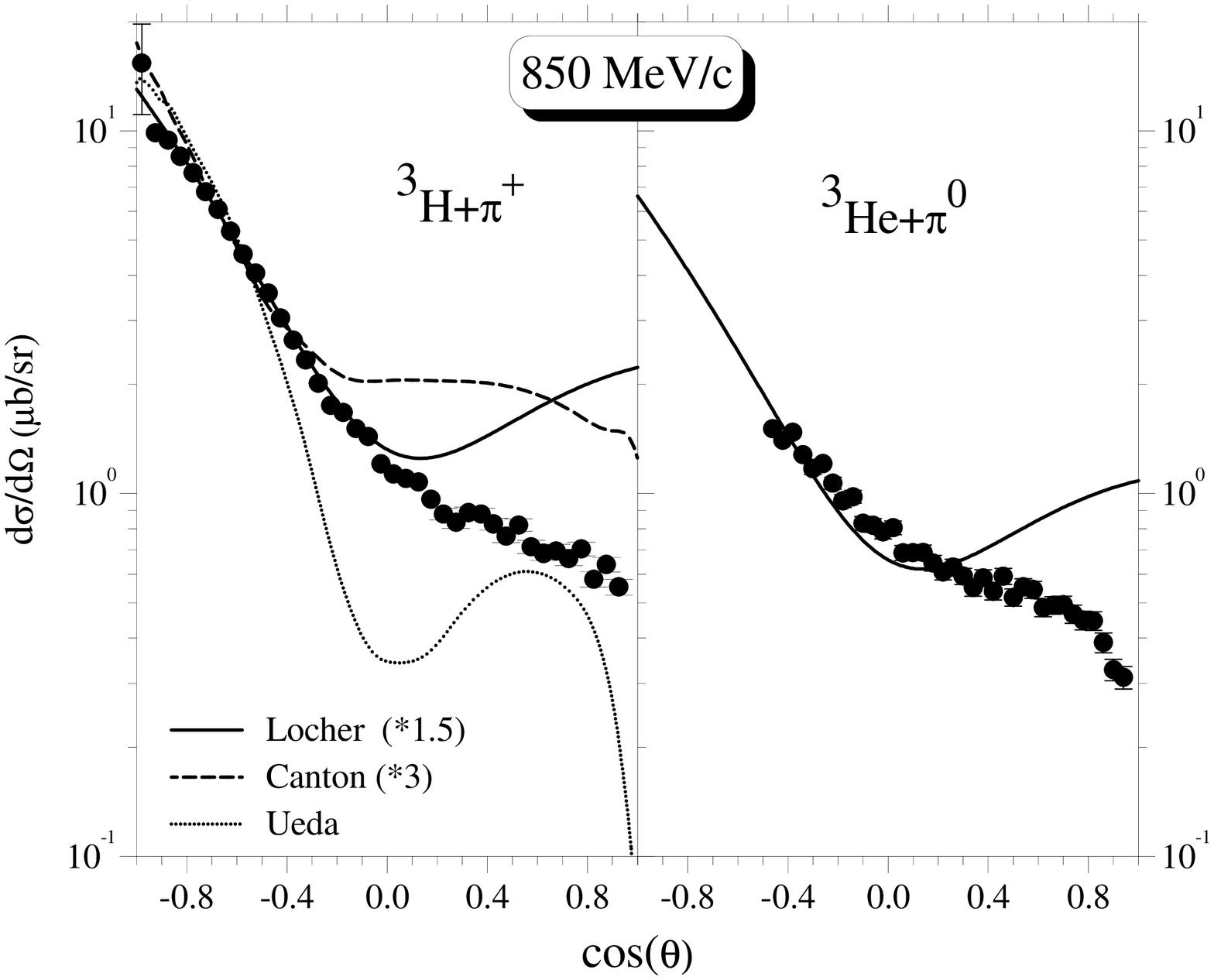,width=14cm} \end{center}
\caption{\label{cross_section_850} Same as Fig. \ref{cross_section_800}
for 850 MeV/c with an additional calculation by Ueda \cite{Ued89}. In
the latter the amplitudes have been normalized by a factor of 0.1.}
\end{figure}

The measured angular distributions in the centre of mass system
are shown in Fig. \ref{cross_section_750},
\ref{cross_section_800}, and \ref{cross_section_850} for beam
momenta of 750 Mev/c, 800 MeV/c, and 850 MeV/c, respectively.
All data are from the measurements with the germanium wall except the
triton zero degree measurement corresponding to $\cos (\theta)\approx
-1$ in the centre of mass system. These points were obtained with the
help of the magnetic spectrograph. Because of the small acceptance left
over for this device by the hole in the germanium wall the count rate
was rather small. An additional uncertainty is due to the size of the
acceptance which may result from a slightly inclined beam together with
the rather thick target and a beam having widths of $\sigma_x=0.66$ mm
and $\sigma_y=0.53$ mm. We assume this additional uncertainty to be
20$\%$ and add it to the count rate error in quadrature. For ${^3He}$
emission at smaller energies for the 750 MeV/c and 800 MeV/c beam
momenta the analysis was based on the almost linear relation between
kinetic energy and emission angle for the two body reaction, as
mentioned above. This method leads to a larger background than in case
of coincidence measurements esp. due to the triton events with not too
different kinematics. Therefore, the statistical error bars are much
larger for these data which are shown as squares in Fig.'s
\ref{cross_section_750} and \ref{cross_section_800}. These points show
some deviations from the points obtained from the coincidence
measurements because of threshold effects. The Quirl had a thickness of
1.3 mm. $^3He$ ions with approximately 70 MeV kinetic energy are
stopped in this detector. This corresponds to $\cos (\theta)\approx
-0.70$ and $\cos (\theta)\approx -0.86$ for the beam momenta of 750
MeV/c and 800 MeV/c, respectively. Ions with slightly larger energy can
fall below thresholds in the trigger electronics or the ADC. This
effect leads to a reduction of counting rate for larger emission angles
and to an increased counting rate in a similar interval below these
angles due to the smaller energy deposit of these ions in the Quirl
detector. (In principle a hole should appear but this is smeared out
due to the longitudinal straggling.) We have studied this effect with
Monte Carlo calculations and found that differential cross sections can
be different up to $11\%$. The data are finally corrected for this
effect.

The shown error bars include the statistical uncertainty as well as the
systematic one from background subtraction. This is larger in the case
of tritium emission than for ${^3He}$ emission while the pure
statistical uncertainty behaves opposite.

The cross sections for both reactions show a backward peaking of the
$A=3$ nuclei which corresponds to a forward peaking of the pion. The
angular distribution for the smallest beam momentum shows an
exponential decrease with increasing $\cos (\theta)$. For the larger
beam momenta an additional component at forward angles shows up which
seems to be less angle dependent. Also shown in Fig.'s
\ref{cross_section_750}, \ref{cross_section_800}, and
\ref{cross_section_850} are model calculations which will be discussed
further down. Only the statistical errors are shown. In addition, there
are systematic uncertainties due to target thickness (5\%), luminosity
calibration (7\%) and detector response and analysis method (3\%). Only
the last contribution is different for the two reactions. The same is
true for the energy loss corrections in the target. This leads to a
total systematic error of up to 9.1$\%$.

\section[Discussion]{\label{Discussion}Discussion}
\subsection[Comparison with other data]
{\label{Data Comparison}Comparison with other data}
\begin{figure}[htbp]
\begin{center}
\epsfig{figure=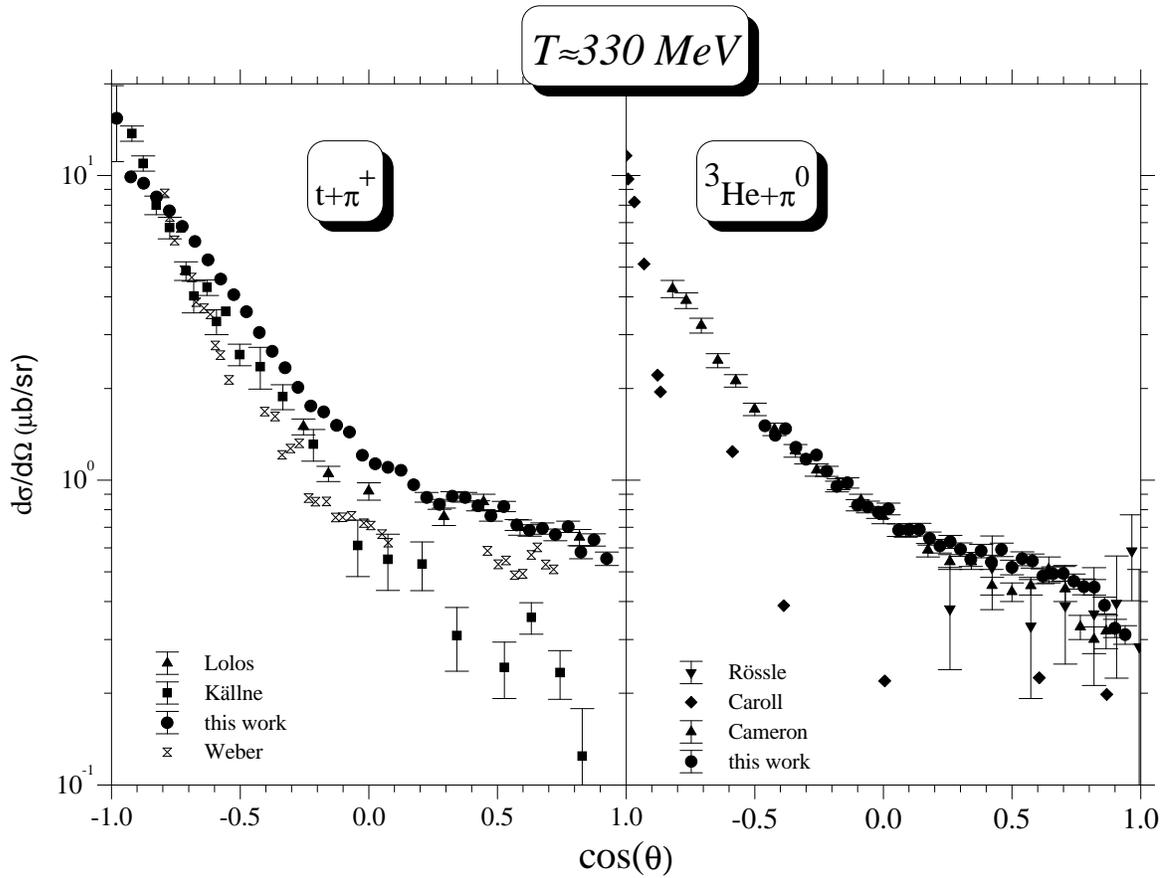, width=18 cm, angle=0}
\end{center}
\caption{\label{data_compare}Comparison of the present data at 850
MeV/c (full dots) with data from Ref.'s  \cite{Kae81,Lol82,Web91} for
the ${^3H}\ {\pi^+}$ reaction (left part) and from
\cite{Dut81,Cam87,Car78} for the ${^3He}\ {\pi^0}$ reaction (right
part).}
\end{figure}

We will first compare the present results with older data. This is
done in Fig. \ref{data_compare}.
The data from Ref. \cite{Kae81} and Ref. \cite{Web91} are from pion
absorption and are transformed assuming charge symmetry as well as time
reversal invariance and hence these data were transformed using
detailed balance. All previous data have larger error bars than the
present ones.  For the charged exit channel, the older data seem to
follow a steeper angular dependence than the present data. The cross
sections from K\"{a}llne et al. \cite{Kae81} for the forward angles are
smaller than the present ones.  For the neutral exit channel there is a
striking agreement between the present data and those of Cameron et al.
\cite{Cam87}, but the data from Carroll et al. \cite{Car78} differ. The
data from R\"{o}ssle at al. \cite{Dut81} are for the related $nd\to t\pi^0$
reaction. They have extremely large error bars. The trend in the latter
data to rise at forward angles is not seen in the present data and
seems to be not statistically significant.

\subsection[Forward-backward and total Cross sections]{\label{For_back}Forward-backward
and total cross sections}

In order to extract total cross sections as well as to extrapolate the
present measurements, analytic functions were fitted to the angular
distributions. A Legendre polynomial of fourth order and fifth order in
the case of 850 MeV/c was found to yield smallest $\chi^2/$degree of
freedom and the final results are from these fits. Also an exponential
with a constant could account for the data, however with much larger
uncertainties. This may have its origin in the different number of
fitted parameters. The total cross section is obtained by integrating
the fitted angular distributions over the full solid angle. Due to the
ambiguity in the order of the Legendre polynomial and the corrections
applied to the data close to the detector acceptance limit a systematic
uncertainty shows up. In case of the ${^3He}$ emission the backward
angle range causes some errors. For the two lower beam momenta the
different method of measurement without coincidence compared to the
forward emitted particles introduced larger error bars and a specific
structure as is discussed above. The influence of this structure on the
total cross section is estimated by simulations. As a result, a
systematic error in addition to the fit error shows up. For 850 MeV/c
no such data exist because of the coincidence requirement in the
hardware trigger. The influence of the missing data was studied by
truncating the corresponding triton angular distribution and the
${^3He}$ distribution from Ref. \cite{Cam87} suggesting a systematic
uncertainty of 10$\%$. The total cross sections for both reactions are
compiled in Table \ref{Tab_1} with the first error the statistical
error and the second error the systematical one as discussed. In
addition, there is a systematic uncertainty of 9.1$\%$ from target
thickness, beam intensity measurements and cuts during the data
evaluation processes. The deduced cross section values increase with
beam momentum for both reactions.

\begin{figure}[htbp]
 \begin{center}
  \epsfig{figure=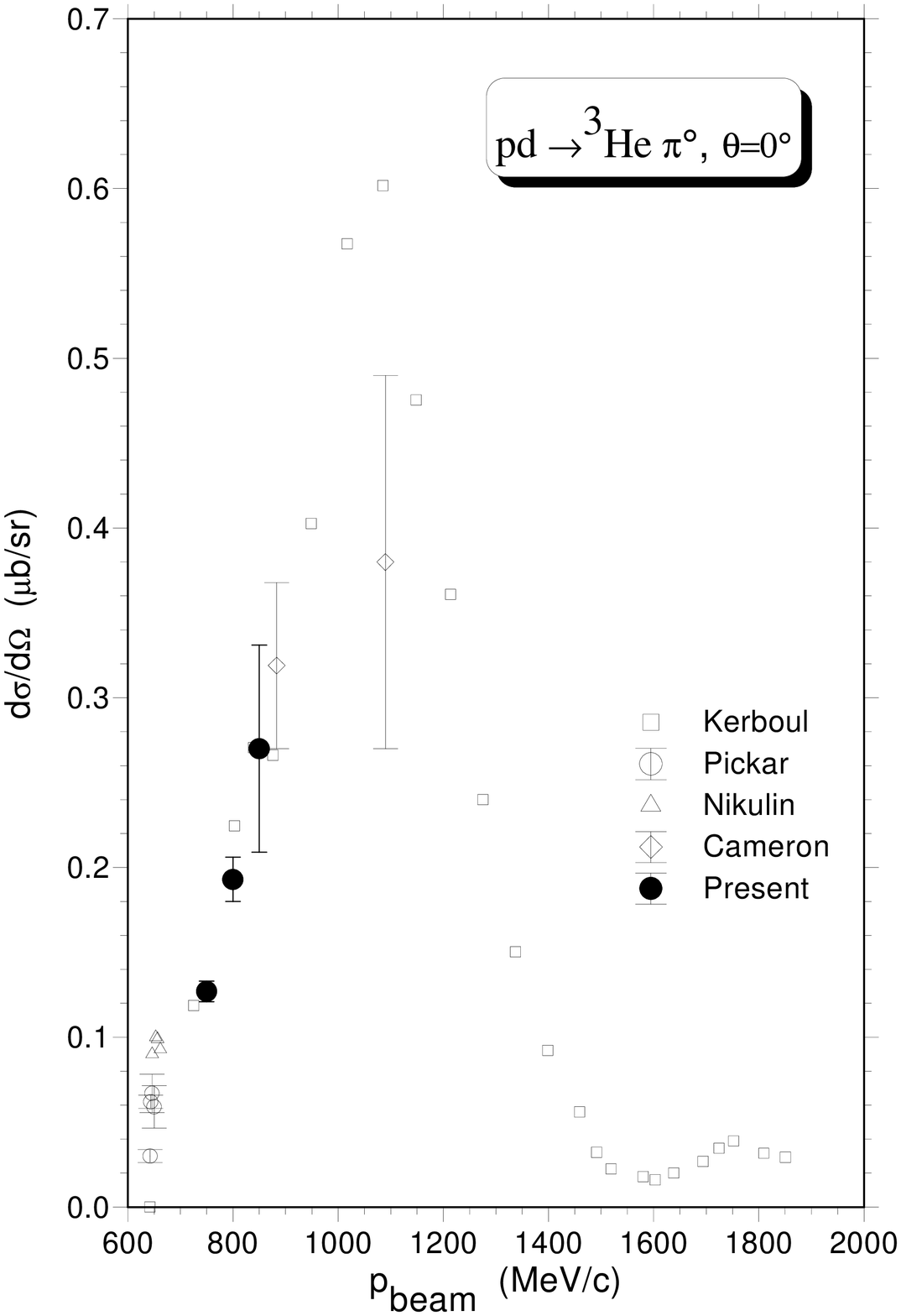, width=6cm, angle=0}
    \hspace{2cm}
   \epsfig{figure=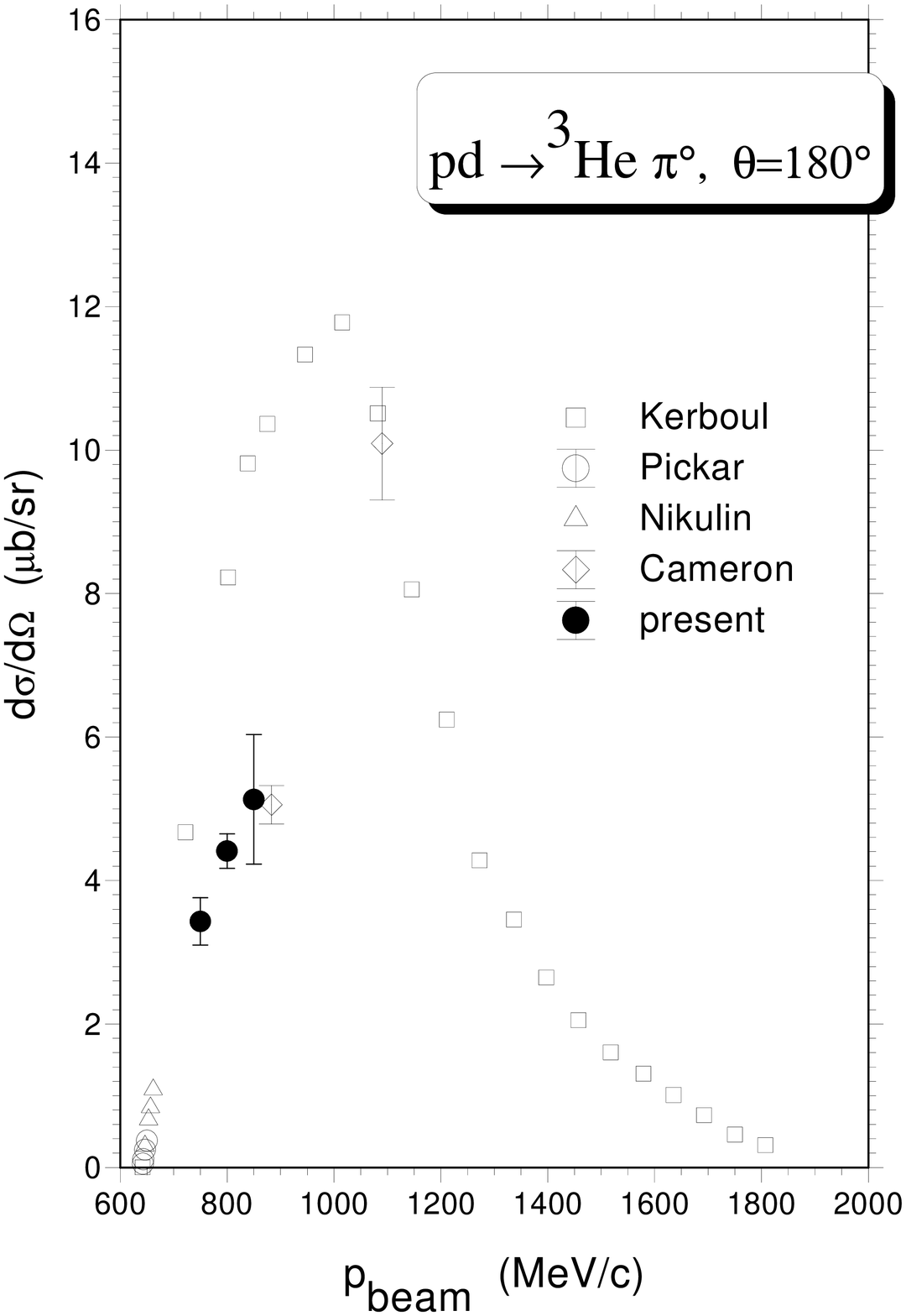, width=6cm, angle=0}
 \end{center}
    \begin{multicols}{2}
\caption{\label{exfu_0_deg} Excitation function for the
$pd\to{^3He}\pi^0$ reaction at a ${^3He}$ emission angle of 0 degree.
The present results are derived by extrapolation of a fit with an
exponential plus a constant to the data. }
\caption{\label{exfu_180_deg}Same as Fig. \ref{exfu_0_deg} but for an
emission angle of the ${^3He}$ of 180 degree. The present results are
derived by extrapolation of the fitted Legendre polynomials (full dots)
including the uncertainties due to different extrapolation methods.}
\end{multicols}
\end{figure}

We proceed by extrapolating to $\cos (\theta ) = - 1$ and $\cos (\theta
) =  +1$ in the same way. The results are shown in Fig.
\ref{exfu_0_deg} and Fig. \ref{exfu_180_deg}, respectively, together
with the results from Ref.'s \cite{Ker86,Pic92,Nik96,Cam87}. Only the
data from Kerboul et al. \cite{Ker86} are direct measurements; all
other data are obtained by us by extrapolation of fitted Legendre
polynomials. For the distribution from Ref. \cite{Cam87} at a momentum
of 1090 MeV/c one point, being off the distribution by almost four
standard deviations, was excluded from the analysis. The best fitting
Legendre polynomial had an order of 6 in this case.
The extrapolation method leads to rather large uncertainties especially
for $\cos(\theta)=1$. Unfortunately, the measurement of the cross
sections for $\cos(\theta)=\pm 1$, which was foreseen to be done with
the magnet spectrometer, failed, because of to thick material in the
focal plane with respect to the large stopping power of ${^3He}$. The
Legendre polynomials have the tendency to underestimate the cross
sections at $\cos(\theta)= 1$. For these cases we show the more robust
results of the exponential fits. They are in nice agreement with the
data from Kerboul et al. \cite{Ker86}. The present backward angle
results seem to be slightly smaller than the measurements by Kerboul et
al. \cite{Ker86}. These large backward cross sections may be the origin
for the too large total cross sections extracted from them  by Pickar
et al. \cite{Pic92} as discussed in the introduction. The present data
agree with the result of Cameron et al. \cite{Cam87} as could be
expected from Fig. \ref{data_compare}

\begin{table}
\caption{\label{Tab_1}Total cross sections for the two reactions as
function of the beam momentum. The first error is the statistical
uncertainty from the Legendre polynomial fit the second the
systematical uncertainty due to the ambiguities in the data at backward
angles and degree of the Legendre polynomial fitted. In addition there
is an overall uncertainty of $\pm 9.1\%$ due to target thickness,
luminosity calibration and analysis method. The uncertainty from the
fit contains the systematical uncertainty resulting from background
subtraction.}
\begin{center}
\begin{tabular}{|c|c|c|}
\hline
 momentum (MeV/c) & $\sigma({^3He}\ {\pi^0})\ (\mu b)$ &
 $\sigma({^3H}\ {\pi^+})\ (\mu b)$ \\
\hline
       750        & $12.48\pm 0.17\pm0.38$ & $25.42\pm 0.30\pm 1.3$\\
       800        & $14.43\pm 0.13\pm0.38$ & $28.05\pm 0.18\pm 0.20$\\
       850        & $15.54\pm 0.30\pm1.55$ & $34.90\pm 0.23\pm 0.05$\\
\hline
\end{tabular}
\end{center}
\end{table}

\begin{figure}[htbp]
 \begin{center}
  \epsfig{figure=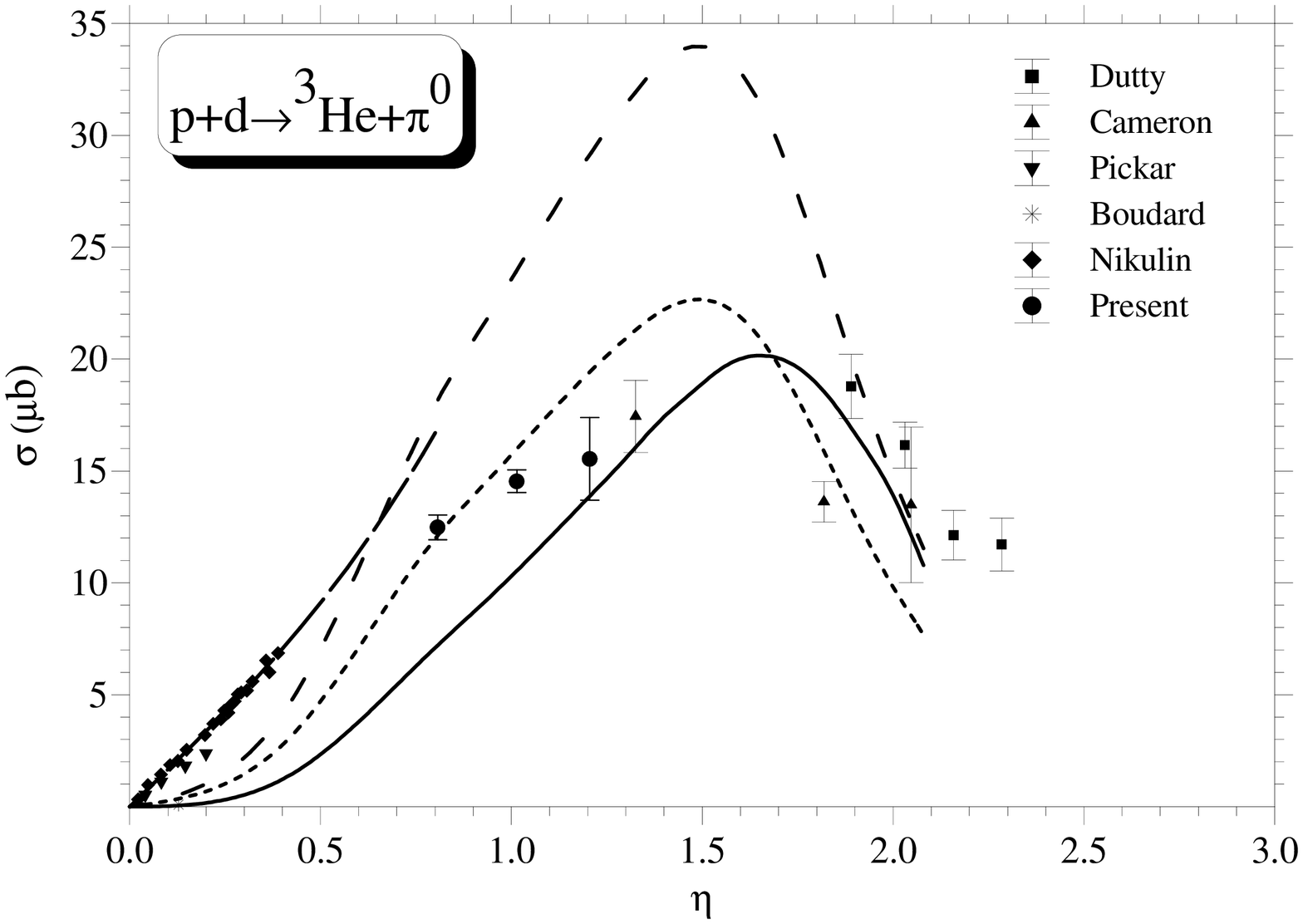, width=12cm, angle=0}
 \end{center}
\caption{\label{exfu_he}Excitation function for the $pd\to {^3He}\pi^0$
reaction. The present data (full dots) are compared with other data
indicated by different symbols. The inner error bars denote the
statistical errors, the outer bars the systematical errors. The long
dashed curve is again the low energy fit to the Germond-Wilkin model,
the solid curve is a calculation in terms of the Locher-Weber model.
The short dashed and dotted curves are similar calculations but using a
different ${^3He}$ form factor and different normalisation (see
subsection \ref{Model}).}
\end{figure}

\begin{figure}
 \begin{center}
  \epsfig{figure=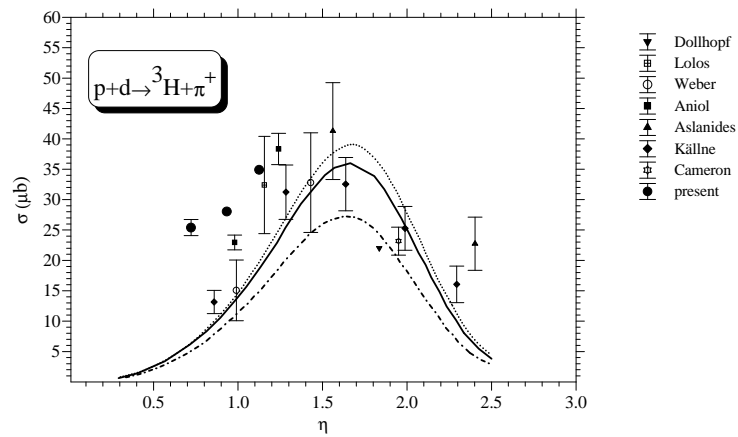, width=12cm, angle=0}
 \end{center}
\caption{\label{exfu_trit} Excitation function for the
$pd\to{^3H}\pi^+$ reaction. The present data are indicated by full
dots, others by different symbols. The statistical errors for the
present data are in the size of the symbols. Also shown is the
systematical error. The data from Ref.'s \cite{Web91,Ani86,Kae81} are
from negative pion absorption on ${^3He}$. The curves are model
calculations \cite{Can98} using the Bonn B potential (solid curve),
Bonn A (dotted curve), and Paris interaction (dashed curve). }
\end{figure}

The total cross sections are compared in Fig.'s \ref{exfu_he} and
\ref{exfu_trit} with data from Ref.'s
\cite{Dut81,Cam87,Bou88,Nik96,Pic92} for the neutral exit channel and
from Ref.'s \cite{Kae81,Cam81,Web91,Ani86,Asl77,Dol73} for the charged
exit channel. In addition to the cases mentioned above, the data from
Aniol et al. \cite{Ani86} and Weber et al. \cite{Web91} were
transformed using detailed balance.
The present cross sections for the $pd\to {^3He}\ {\pi^0}$ reaction
seem to nicely interpolate between the near threshold data and those in
the resonance region. The calculation employing the low energy
parameters of the Germond--Wilkin model \cite{Ger90} shown as dashed
curve and previously discussed in the introduction seem to come close
to the present data point at 750 MeV/c bombarding momentum. Also shown
is a calculation in the framework of the Locher--Weber prescription
\cite{Loc74}, which will be discussed below.

For the $pd\to {^3H}\ {\pi^+}$ reaction, there are a few points in the
range of the present data. Almost all of them stem from studies of
negative pion absorption transformed by detailed balance assuming
charge symmetry and time--reversal invariance. This disagreement was
already noted by the authors of the  latter data \cite{Ani86}.

\subsection[Model Comparison]{\label{Model}Model
comparison}

As discussed in the introduction, there are numerous models for the
mechanism of these reactions. Here, we restrict ourselves to
comparisons with published calculations for beam momenta close to the
present ones or perform such calculations in the very transparent
Locher--Weber model \cite{Loc74}. The differential cross section in
this model is given by
\begin{equation} \label{Locher}
\frac{{d\sigma }}{{d\Omega }}\, = \,S\,\,K\,|F_D (q) - F_E (q)|^2
\,\frac{{d\sigma }}{{d\Omega }}(pp \to d\pi ^ +  )
\end{equation}
with $S$, a spin factor, $K$ a kinematical factor, and $F_D$, the
direct form factor and $F_E$, the exchange form factor, i.e. an elastic
$\pi d$ scattering after pion emission from the incident proton. The
form factors were evaluated with emphasis on the short--range
components of the deuteron and the triton. This is achieved by fitting
the free parameters in a Hulth\'{e}n function to the deuteron--charge form
factor and similarly for tritium, the parameters for the Eckart
function and the 3-pole function to the tritium--charge form factor.
These form factors are structure functions obtained from elastic
electron scattering and should not be mixed with its component the
monopole charge form factor, having a similar shape. The latter is the
Fourier transform of a monopole density containing both the s-state and
d-state deuteron wave functions times the sum of the proton and neutron
form factors. The figure captions in Ref. \cite{Loc74} are somewhat
confusing in this respect. It should be mentioned that the contribution
of the magnetic dipole form factor to this quantity is very small.
Since the experimental input into the fitting functions is rather old,
we have compared these functions with newer measurements and find nice
agreement for the deuteron form factor with results from Platchkov et
al. \cite{Pla90} and for the ${^3He}$ form factor with those from
Amroun et al. \cite{Amr94}. However, recent data from JLAB
\cite{Ale99,Abb99} show that the assumed Hulth\'{e}n function is too small
for large momentum transfers.

We have fitted Legendre polynomials to the cross section of the
elementary reaction including new data in the threshold region
\cite{Dro98}. The energy dependence of the Legendre coefficients was
fitted in 3 different but overlapping regions: threshold region,
resonance region and above, although that interval is presently not of
interest. Since the model does not treat the internal structure of the
nuclei nor angular momenta it can not predict spin observables. One
model ambiguity is the choice of the momentum of the struck nucleon.
Since ignoring the Fermi motion in the deuteron leads to a wrong
resonance position, the second option of Ref. \cite{Loc74} was chosen,
i. e. ignoring of the Fermi motion in the triton. The emission angle
was always assumed to be the same for both reactions.

Calculations employing the 3-pole wave function were found to be
always smaller in the exponential part than those using the Eckart
wave function. Here we present results only for the Eckart wave
function. Such calculations are shown in Fig.'s
\ref{cross_section_750}, \ref{cross_section_800}, and
\ref{cross_section_850}. A normalization factor of 1.5 was always
applied. For ${^3He}$ emission, just an isospin factor of 0.5 was
used. The small differences result from the different kinematics.

At 750 MeV/c, the data are larger than the calculations, whereas for
the higher beam momenta the part of the angular distribution showing an
exponential dependence is nicely reproduced. The strong increase in the
calculations for forward angles is not supported by the data. This
increase is reduced if direct and exchange contributions are added
incoherently. The total cross section prediction within this model is
shown in Fig. \ref{exfu_he}. It seems to work well in the resonance
region, but fails in the threshold region.

In order to study the momentum range being sensitive to the present
reactions we have introduced a step function in the integrals of the
form factors. The results become almost independent if the truncation
omits the range larger than 4$fm^{-1}$, However, in this range the
quadrupole form factor is already larger than the monopole form factor,
thus making the D-state important which is missing in the present
model.

 In order to study the sensitivity of the model calculations
to the form factors we have chosen the options used by Fearing
\cite{Fea77}: again the Hulth\'{e}n form for the deuteron but exponential,
Gaussian, and Irving--Gunn forms for the ${^3He}$ using the range
parameters from ref. \cite{Fea77} but the same normalization as Locher
and Weber. In these calculations the same pion centre of mass momentum
in the elementary as well as the $p+d$ reaction was assumed.
Exponential and Gaussian give almost the same results as the Eckart
function. The Irving--Gunn function leads to an almost symmetric
angular distribution. Such a behaviour is in disagreement with the
data. It is only this function which yields a different shape. The
three others are very close to each other in the range up to 4 fm
except for the Eckart function, which drops down for radii below the
maximum at 1 fm. The resulting excitation function for the choice of an
exponential form is also shown in Fig. \ref{exfu_he}. We have applied
two normalizations: one to fit the present data and the same curve
multiplied by 1.5 to investigate the near threshold region. The maximum
is shifted towards smaller beam momenta when compared to the
calculation employing the Eckart form. This is due to the assumptions
about the effective energy and momentum in the $pp\to d\pi^+$
subamplitude. The calculation with larger normalization yields larger
cross sections than the Eckart form. However, the resulting cross
section values are below the experimental points. This is similar to
the results obtained by Falk \cite{Falk}. His model is based on the
same physical picture but more refined to also predict spin
observables. The calculation underestimates the cross section data
close to threshold  by a factor of two, but overestimates data in the
resonance region by a factor of approximately 2.6.

Canton and Schadow \cite{Can98} performed much more rigorous
calculations; again the $pp\to d\pi^+$ reaction is the underlying
mechanism. The three body wave function was calculated from
nucleon--nucleon potentials employing three body calculations. In
Fig.'s \ref{cross_section_800} and \ref{cross_section_850} the
calculations are compared to the data. A normalization constant of 3 is
used. The exponential behaviour at backward angles is reproduced.
Forward angles are overestimated. Total cross sections as a function of
the previously defined quantity $\eta$ are shown in Fig.
\ref{exfu_trit} together with data. The calculations were performed for
different nucleon--nucleon potentials: Bonn A, Bonn B, and Paris
potential. The quality of the data in the resonance region does not
allow discrimination between different potentials. It is surprising
that in the range below $\eta=1$ all calculations with different
potential choices start to merge whereas it is just this range where
there are large differences in the calculations for the $pp\to d \pi^+$
reaction \cite{Can98}. It is interesting to note that the calculations
for this reaction agree only with the experimental data when a strong
final state between the deuteron and the pion is included in the
calculation. The inclusion of this final state interaction increases
the cross section by typically an order of magnitude which is
surprising with respect to other calculations. This model also fails to
reproduce the large differential cross sections shown in Figure
\ref{exfu_180_deg}.

Ueda \cite{Ued89} attacked the problem by splitting the many--body
process into coupled multi-, three-, and two-body systems which were
treated in a relativistic and unitary approach. Numerical input is
obtained by adjusting potential parameters to two body scattering
amplitudes: $NN-NN$, $NN-d\pi$, $\pi d-\pi d$, $NN-N\Delta$, $\pi
d-N\Delta$, and $N\Delta -N\Delta$. A calculation without further
normalization is shown in Fig. \ref{cross_section_850}. It should be
mentioned that a damping factor of 0.1 was applied to the otherwise too
large amplitudes. The large cross sections at backward angles are
reproduced. The oscillations at forward angles are not supported by the
data. The success of this model may be that a lot of different graphs
like $NN$ and $N\Delta$ correlations (see Ref. \cite{Saino}), multiple
scattering effects (Ref. \cite{Fea77}) and three nucleon mechanism
(Ref. \cite{Laget}) are automatically included in this approach.

\section[Isospin Symmetry]{\label{Isospin}Isospin Symmetry}

It was already pointed out by Ruderman \cite{Rud52} that data of the
present type, i. e. differential cross sections for the two isospin
related reactions, should allow for testing the validity of isospin
symmetry. However, the two isospin related reactions have different
Q-values leading to differences even for the same beam momentum.The
ratios for the total cross sections yield $1.97\pm 0.05$ for the three
measurements (see Table 1). We compare the two exit channels on the
level of differential cross section. The mean ratios are $1.89\pm0.04$,
$1.87\pm0.03$ and $1.64\pm0.04$ for the beam momenta of 750 MeV/c, 800
MeV/c and 850 MeV/c, respectively. However, the measured ratios are not
constant, they show a trend to decrease with increasing emission angle.
This may point to different shapes of the angular distributions for the
two reactions. Already the different pion masses lead to trivial
deviations from 2 for the ratio Eq. \ref{mean}. To take this trivial
deviation in some approximate way into account, we study the ratio for
the differential cross sections instead at the same $p^{cm}_\pi$ and
the same $\eta$--value. The angular distributions for the
$p+d\rightarrow {^3H}+\pi^+$ reaction at the same $p^{cm}_\pi$ and
$\eta$-value are obtained by linear interpolating the angular
distributions for the neighbouring $p^{cm}_\pi$ and $\eta$-values,
respectively. The principal advantage of cancellation of systematical
errors is lost in part by doing so and the linear interpolation may
introduce another systematical error in the percent range. The results
are $1.91\pm 0.04$ for $p^{cm}_\pi=108.8\ MeV/c$ and $1.97\pm 0.03$ for
$p^{cm}_\pi=136.6\ MeV/c$.
\begin{figure}[htb]
\begin{center}
\epsfig{file=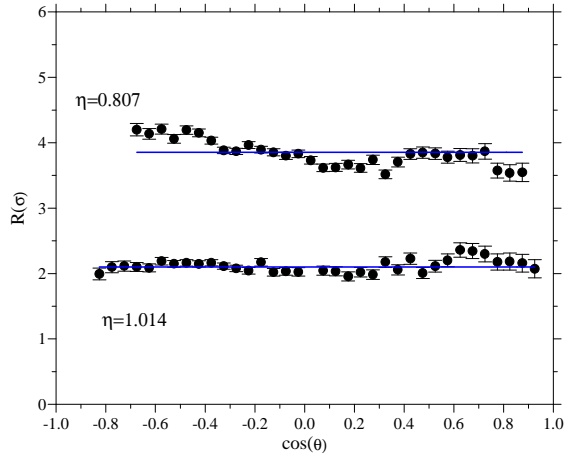, width=10cm}
\end{center}
\caption{\label{xsratio}The ratio of the differential cross sections
for the two reactions as a function of the emission angles in the
centre of mass system for the indicated values of $\eta$. A constant of
2 was added to the upper data set.}
\end{figure}
The ratios for the two reactions obtained for constant $\eta$-values
are shown in Fig. \ref{xsratio}. This procedure yields for $\eta =
0.807$, $1.905\pm0.034$ and for $\eta =1.014$, a ratio $2.145\pm
0.013$. In all cases we have corrected the $^3H+\pi^+$ data for the
Coulomb effects by the usual Gamow factor being 2 to 3\% depending on
the momentum.  All three methods lead to a weighted mean value close to
two. However, it is not clear on which level one should compare. In
addition, trivial effects due to different pion masses or pion or beam
momenta remain. We may, therefore, compare matrix elements instead of
differential cross sections. They are calculated according to
\begin{equation}\label{matrix}
\frac{{d\sigma (\theta )}}{{d\Omega }}\, = \,\frac{{(2s_b  + 1)(2s_B  +
1)}}{{(2\pi)^2\hbar^4}}\frac{{p_b^{cm} }}{{p_a^{cm} }}\frac{{W_a W_b
W_A W_B }}{s}|M(t)|^2
\end{equation}
for a reaction $a+A\to b+B$. In Eq. \ref{matrix} $s_i$ denotes the
spins, $p_i^{cm}$ the momenta of the initial and final state in the c.
m. system, $W_i$ the total energies and $s$ the total energy squared.
The matrix element $|M|$ is a function of the four momentum squared
$t=t[\cos(\theta)]$ and is measured in fm$^3$MeV. Eq. \ref{matrix} is
the relativistic formulation of the well known non-relativistic result
(see Eq. 4.8 in Ref. \cite{Per87}).

Such elimination of the phase--space factor from the data  should lead
to a more rigourous result as was pointed out by Silverman et al.
\cite{Sil85}. The ratio of the matrix elements squared are $2.06\pm
0.06$, $2.02\pm 0.03$ and $1.73\pm 0.04$ for the same total energy $s$,
$1.95\pm 0.04$ and $1.98\pm 0.04$ for the same $p^{cm}_\pi$, and
$1.87\pm 0.03$ and $2.08\pm 0.01$ for the same $\eta$-values. It is
interesting to note that the correction is smallest (less than 1\%) for
constant $p^{cm}_\pi$. It is always the measurement at 850 MeV/c which
has a ratio much smaller than 2. The influence of this measurement on
the results for constant $p^{cm}_\pi$ or constant $\eta$ is small. If
one neglects this one ratio, the mean values are $2.03\pm 0.02$,
$1.97\pm 0.03$ and $2.02\pm 0.07$ for constant $s$, $p^{cm}_\pi$ and
$\eta$, respectively. Which criterion is the best is not clear. The
latter two methods suffer from different measurements and the
interpolation procedure resulting in some systematic error being at
least of the order of the statistical error.

\begin{figure}[htbp]
\begin{center}
\epsfig{file=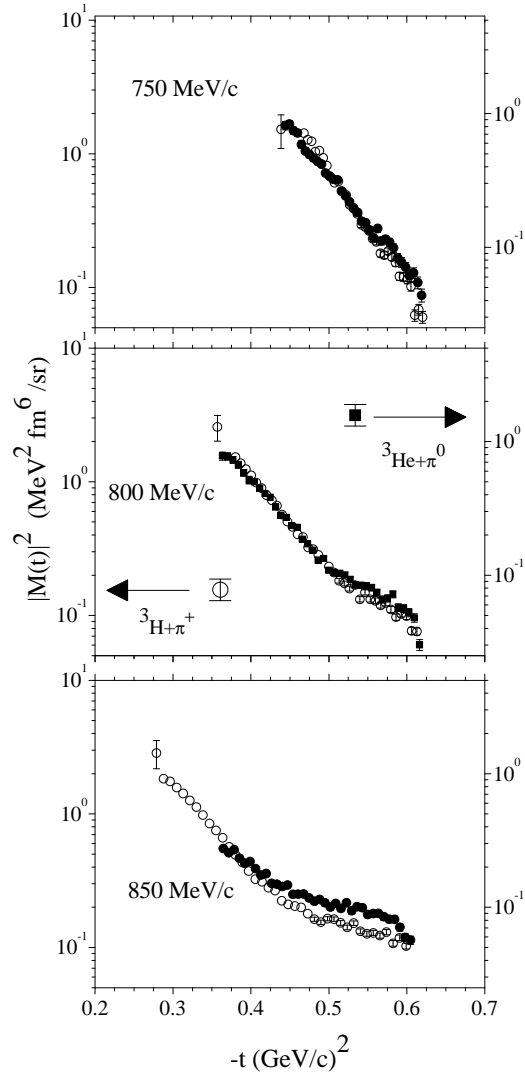,height=15cm} \end{center}
\caption{\label{matrix_compare} The deduced squared matrix elements for
the two reactions and for the three bombarding energies. The y-axes for
the $pd\to {^3H}\ {\pi^+}$ reactions are on the left side while those
for the $pd\to {^3He}\ {\pi^0}$ reactions are on the right side. For
each beam momentum, the two axes are adjusted by a factor of two in
order to simulate the naive isospin ratio.}
\end{figure}

One may inspect the matrix elements directly. This is done in Fig.
\ref{matrix_compare}.
The matrix elements for the lowest beam momentum show an almost
exponential dependence on the four--momentum transfer squared. For the
highest momentum, an additional component for large--momentum transfer
shows up, the same as the differential cross sections. The data
definitely show a dependence on the beam momenta. This is in contrast
to the conjecture by Silverman et al. \cite{Sil85} that the invariant
matrix element squared is independent of the beam energy which, by the
way, is not supported by any theory.

The y-axis for each beam momentum are adjusted to each other for the
two reactions by a factor of two, in order to get an estimate of the
validity of isospin symmetry. There is a common tendency in the data:
the matrix element for the ${^3He}+\pi^0$ reaction is smaller than the
isospin corrected one for the ${^3H}+\pi^+$ reaction for small momentum
transfer and larger for large momentum transfer. This behaviour becomes
more pronounced with increasing beam momentum. This may have its origin
in different slopes for different $s$ and $t$, the data decrease with
increasing pion momentum. This leads to a t-dependence of the ratio
\begin{equation}\label{mean}
 R\ =\ \frac{|M(p+d\rightarrow {^3H+}\ \pi^+,t)|^2}{|M(p+d\rightarrow
  {^3He}+\pi^0,t)|^2}
\end{equation}
for data obtained at the same beam momentum.
\begin{figure}[htbp]
\begin{center}
\epsfig{figure=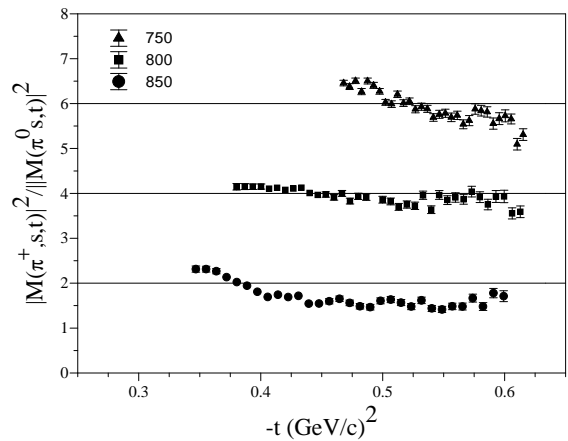,width=10cm, angle=0}
\end{center}
\caption{\label{ratio_mat}The ratios of the matrix elements squared for
the three beam momenta. A constant of 4 is added to the results for 750
MeV/c and 2 for those at 800 MeV/c.}
\end{figure}
The ratios, not Coulomb corrected, show an decrease with increasing
momentum transfer which may have its origin in different kinematics.
The ratio for 850 MeV/c is flat for high momentum transfers and in this
region much smaller than two. Whether these effects result from
interferences with Coulomb effects or are due to increasing importance
of $\Delta$ resonance excitation with increasing bombarding energy
needs further studies.

\section[Conclusion]{\label{Conclusion}Conclusion}
In summary, we have measured simultaneously the two reactions
$p+d\rightarrow {^3H}+\pi^+$ and $p+d\rightarrow {^3He}+\pi^0$ in the
intermediate region between the near threshold and the resonance range.
A liquid deuterium target with very thin walls reduced empty--target
corrections to a negligible level. By the present method problems due
to normalization to beam current, target thicknesses, solid angle, dead
time corrections etc. are avoided. It is the first time that the ratio
of the two reactions is studied over a large angular range. The
differential cross sections as well as the total cross sections for the
$p+d\rightarrow {^3He}+\pi^0$ reaction bridge the previous gap between
data in the threshold region and the resonance region. However, for the
$p+d\rightarrow {^3H}+\pi^+$ reaction, a discrepancy with data for the
time--reversed reaction shows up. This was already mentioned by Aniol
et al. \cite{Ani86} and may have its origin in the differences between
initial and final--states due to Coulomb effects in the two reactions.

It is found that models based on the input of $p+p\rightarrow d+ \pi^+$
cross sections or amplitudes yield total cross sections for the $p+d$
reactions which are close to $p+p\rightarrow d+ \pi^+$ cross sections
scaled down by factors of 80 and 160 for the $p+d\rightarrow
{^3H}+\pi^+$ reaction and $p+d\rightarrow {^3He}+\pi^0$ reaction,
respectively. Since these data follow a different energy dependence
than the present data, the calculations do the same. The enhancement of
the calculated differential cross sections for large momentum transfer
is not supported by the present data. It seems that none of the models
discussed above is able to account for the experimental data over a
large energy range on an absolute base. It is worthwhile to mention
that the S-wave model of Locher and Weber \cite{Loc74} agrees best with
the data in a range where the D-wave in the deuteron contributes most
strongly.

The deduced isospin ratio of $1.64$ up to $1.89$ for the differential
cross sections is smaller than the one found by Silverman et al.
\cite{Sil85} of $2.36\pm 0.11$. The measured ratio of the matrix
elements squared for the two lower beam momenta of $2.03\pm 0.02$ is in
agreement with the result $R=2.13\pm0.13$ obtained by Harting et al.
\cite{Har60} at a beam energy of 591 MeV. At this energy the
differences in the kinematics are expected to be small. K\"{o}hler
\cite{Koh60} estimated that, if the effect of the Coulomb force in the
$^3He$ wavefunction is reduced, the ratio changes from 2 to $2.14 \pm
0.02$ for 600 MeV. Introducing such an additional correction leaves
almost no room for isospin symmetry breaking. Such calculations for
different beam momenta and with modern wavefunctions will be helpful in
answering the question on the sensitivity of isospin symmetry breaking
in the present two reactions. Further measurements will improve the
present error bars.

 \section{Acknowledgement}
We are grateful to the COSY operation crew for their efforts making a
good beam. Support by BMBF Germany (06 MS 568 I TP4), Internationales
B\"{u}ro des BMBF (X081.24 and 211.6), SCSR Poland (2P302 025 and 2P03B 88
08), and COSY J\"{u}lich is gratefully acknowledged.

\end{sloppypar}

\begin{thebibliography}{99}
\bibitem{Rud52} M Ruderman, Phys. Rev. {\bf 87} (1952) 383
\bibitem{Can98} L. Canton, G. Cattapan, G. Pisent, W. Schadow, J. P.
Svenne, Phys. Rev. {\bf C 57} (1998) 1588; L. Canton and W. Schadow,
Phys. Rev: {\bf C 56} (1997) 1231
\bibitem{Dro98} M. Drochner et al., Nucl. Phys.  {\bf A 643} (1998)
55
\bibitem{Ger90} J.-F. Germond and C. Wilkin, J. Phys. {\bf G 16} (1990)
381
\bibitem{Loc74} M. P. Locher and H. J. Weber, Nucl. Phys. {\bf B 76}
(1974) 400
\bibitem{Ker86} C. Kerboul et al., Phys. Lett. {\bf B 181} (1986) 28
\bibitem{Nik96}V. N. Nikulin et al., Phys. Rev. {\bf C 54} (1996) 1732
\bibitem{Bou88} A. Boudard et al., Phys. Lett. {\bf B214} (1988) 6
\bibitem{Cam87} J. M. Cameron et al., Nucl. Phys. {\bf A 472} (1987)718
\bibitem{Dut81} M. Dutty, Diploma thesis, Freiburg 1981 and E. R\"{o}ssle et al., Proc. Conf. on Pion Production and Absorption in Nuclei
(Edt. R. D. Bent), AIP Conf. Proc. {\bf 79} (1982)171
\bibitem{Pic92} M. A. Pickar et al., Phys. Rev. {\bf C 46} (1992) 397
\bibitem{Ber98} A. M. Bernstein, Phys. Lett. {\bf B442} (1998) 20
and references therein
\bibitem{Fea77} H. W. Fearing, Phys. Rev. {\bf C 16} (1977) 313
and references therein
\bibitem{Cre60} A. V. Crewe et al., Phys. Rev. {\bf 118} (1960)
1091
\bibitem{Har60}D. Harting et al., Phys. Rev. {\bf 119} (1960) 1716
\bibitem{Low81} J. W. Low et al., Phys. Rev. {\bf C 23} (1981)
1656
\bibitem{Sil85} B. H. Silverman et al., Nucl. Phys. {\bf A 444} (1985)
621
\bibitem{Cam81} J. M. Cameron et al., Phys. Lett. {\bf 103B}
(1981) 317

\bibitem{Koh60} H. S. K\"{o}hler, Phys. Rev. {\bf 118} (1969) 1345

\bibitem{Jae94} V. Jaeckle, K. Kilian, H. Machner, Ch. Nake, W. Oelert, P.
Turek, Nucl. Instruments and Methods in Physics Research A 349
(1994) 15
\bibitem{Bet99} M. Betigeri et al., Nuclear Instr. Methods in Physics
Research {\bf421} (1999) 447

\bibitem{Gar99} W. Garske, Ph. D. Thesis, Universit\"{a}t M\"{u}nster (2000)
\bibitem{Mac99} H. Machner et al., Nucl. Instr. Methods for Physics Research
{\bf A 437} (1999) 419
\bibitem{Ued89} T. Ueda, Nucl. Phys. {\bf A 505} (1989) 610
\bibitem{Kae81} J. K\"{a}llne, J. E. Bolger, M. J. Devereaux, S. L.
Verbeck, Phys. Rev. {\bf 24} (1981) 1102
\bibitem{Lol82} G. J. Lolos et al., Nucl. Phys. {\bf 386} (1982) 477
\bibitem{Web91} P. Weber et al., Nucl. Phys. {\bf A534} (1991) 541
\bibitem{Car78} J. Caroll et al., Nucl. Phys. {\bf A305} (1978) 502
\bibitem{Ani86} K. A. Aniol et al., Phys. Rev. {\bf C 33} (1986) 1714
\bibitem{Asl77} E. Aslanides et al., Phys. Rev. {\bf 39} (1977) 1654
\bibitem{Dol73} W. Dollhopf et al., Nucl. Phys. {\bf A 217} (1977) 381
\bibitem{Pla90} S. Platchkov et al., Nucl. Phys. {\bf A 510} (1990) 740
\bibitem{Amr94} A. Amroun et al., Nucl. Phys. {\bf A 579} (1994) 596
\bibitem{Ale99} L. C. Alexa et al., Phys. Rev. Lett. {\bf 82} (1999)
1374
\bibitem{Abb99} D. Abbott et al., Phys. Rev. Lett. {\bf 82} (1999) 1379
\bibitem{Falk} W. R. Falk, Phys. Rev. {\bf C 50} (1994) 1574 and {\bf C
61} 034005
\bibitem{Saino} A. M. Green and M. E. Sainio, Nucl. Phys. {\bf A 329}
(1979) 477
\bibitem{Laget} J. M. Laget and J. F. Lecolley, Phys. Lett. {\bf B 194}
(1987) 177
\bibitem{Per87} D. H. Perkins, Introduction to High Energy Physics,
Addison--WesleyPubl. Company 1987






\end{thebibliography}
\end{document}